\pdfoutput=1
\RequirePackage{color}
\RequirePackage{ifpdf}

\documentclass{JINST}

\usepackage{color}

\title{Breakdown voltage of metal-oxide resistors in liquid argon}

\author{L.F. Bagby$^a$, S. Gollapinni$^b$, C.C. James$^a$, B.J.P. Jones$^c$, H. Jostlein$^a$, S. Lockwitz$^a$, D. Naples$^d$, J.L. Raaf$^a$, R. Rameika$^a$, A. Schukraft$^a$, T. Strauss$^e$\thanks{Corresponding author.}, M.S. Weber$^e$ ~and S.A. Wolbers$^a$\\
\llap{$^a$}Fermi National Accelerator Laboratory, PO Box 500, Batavia IL 60510, USA\\
\llap{$^b$}Department of Physics, Kansas State University, 116 Cardwell Hall, Manhattan, Kansas 66506, USA\\
\llap{$^c$}Massachusetts Institute of Technology, 77 Massachusetts Avenue, Cambridge, MA 02139, USA
\llap{$^d$}University of Pittsburgh, Dept. of Physics, 100 Allen Hall, Pittsburgh PA 15260, USA\\
\llap{$^e$}University Bern, LHEP, Sidlerstrasse 5, CH-3012 Bern, Switzerland\\
E-mail: \email{thomas.strauss@lhep.unibe.ch}}

\abstract{We characterized a sample of metal-oxide resistors and measured their breakdown voltage in liquid argon by applying high voltage (HV) pulses over a 3 second period. This test mimics the situation in a HV-divider chain when a breakdown occurs and the voltage across resistors rapidly rise from the static value to much higher values. All resistors had higher breakdown voltages in liquid argon than their vendor ratings in air at room temperature. Failure modes range from full destruction to coating damage. In cases where breakdown was not catastrophic, subsequent breakdown voltages were lower in subsequent measuring runs. One resistor type withstands 131\,kV pulses, the limit of the test setup.}

\DeclareGraphicsExtensions{.pdf,.png,.jpg}

\usepackage{lineno}
\begin{document}
Metal-oxide resistors are chosen for their stable resistances over large temperature ranges and their proven technology for use in high voltage (HV) applications like HV dividers, measuring resistances or variable-frequency drives. Simulations of breakdowns in the unprotected HV divider chain of the MicroBooNE experiment showed that the over-voltage following a discharge is held for a time of 0.1 - 1 s \cite{ref:surge}. This time scale is dictated by the resistances and capacitances of the TPC, rather than the details of spark evolution. With the test setup that was used a HV transient of about ~3\,s could be applied. We characterize different metal-oxide resistors, selected  to cover a broad range of different shapes and coatings,  for their resistances and breakdown voltages in liquid argon. The measurements were performed with a HV test stand at Fermilab. Depending on the maximum applied power load, the resistors were identified as candidates for application in a HV divider chain in a liquid argon time projection chamber, where installing reliable resistors is crucial because access to installed components will be difficult or impossible. 

\section{Breakdown Voltage}
\subsection{Test apparatus}
\begin{figure}[htb]
\centering
\includegraphics[width=1\textwidth,trim={1cm 7.5cm 1cm 4cm},clip]{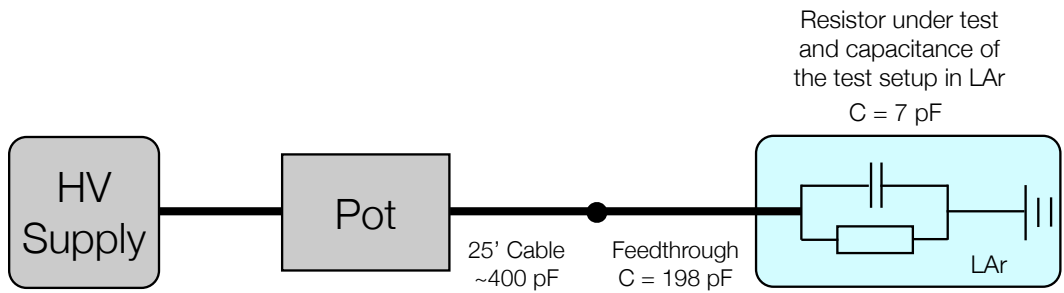}
\caption{Schematic of the test setup.  The section in liquid argon is highlighted in the box on the right.  The test apparatus has an inherent capacitance and is explicitly shown in the diagram.  Only the energy stored downstream of the filter pot is relevant for discharge conditions; the upstream component dissipates slowly to ground in a resistor in the filter pot.}
\label{fig:schematic}
\end{figure}

To test the resistors in liquid argon under high voltage, a test setup was constructed as shown schematically in Figure \ref{fig:schematic}. The setup consists of a HV power supply, high voltage cables, a low pass filter pot, a custom build HV feedthrough, and a test apparatus.

The power supply is a Glassman LX150N12 capable of producing up to $-150$\,kV with a limiting current of 12\,mA (1000~W maximum total power output). For the measurements, the current limit was set to 1\,mA. The cable from the supply to the pot was provided by Glassman and is type DS 2121.

The filter pot consists of a series of eight (TRW MVH-4) resistors immersed in transformer oil (Shell Diala AX). Combined with the capacitance of the cable, it acts as a low-pass filter reducing any high-frequency ripple from the power supply. The filter pot also partitions the energy stored in the system between the power supply and feedthrough, thus reducing the energy impulse sustained during a discharge.

From the filter pot, a 25~foot Dielectric Sciences cable (model 2134) connects to the HV feedthrough. This feedthrough is custom made to apply high voltage to a device within the liquid argon. Here, the feedthrough is a central stainless steel conductor inside of a ultra high molecular weight polyethylene tube (UHMW PE, 1 inch inner diameter; 2 inch outer diameter) that is surrounded by a stainless steel grounding tube that extends into the liquid. The center conductor of the feedthrough contacts the test apparatus to provide the high voltage \cite{ref:docdbHV}.

\begin{figure}[htb]
\centering
\includegraphics[width=0.7\textwidth]{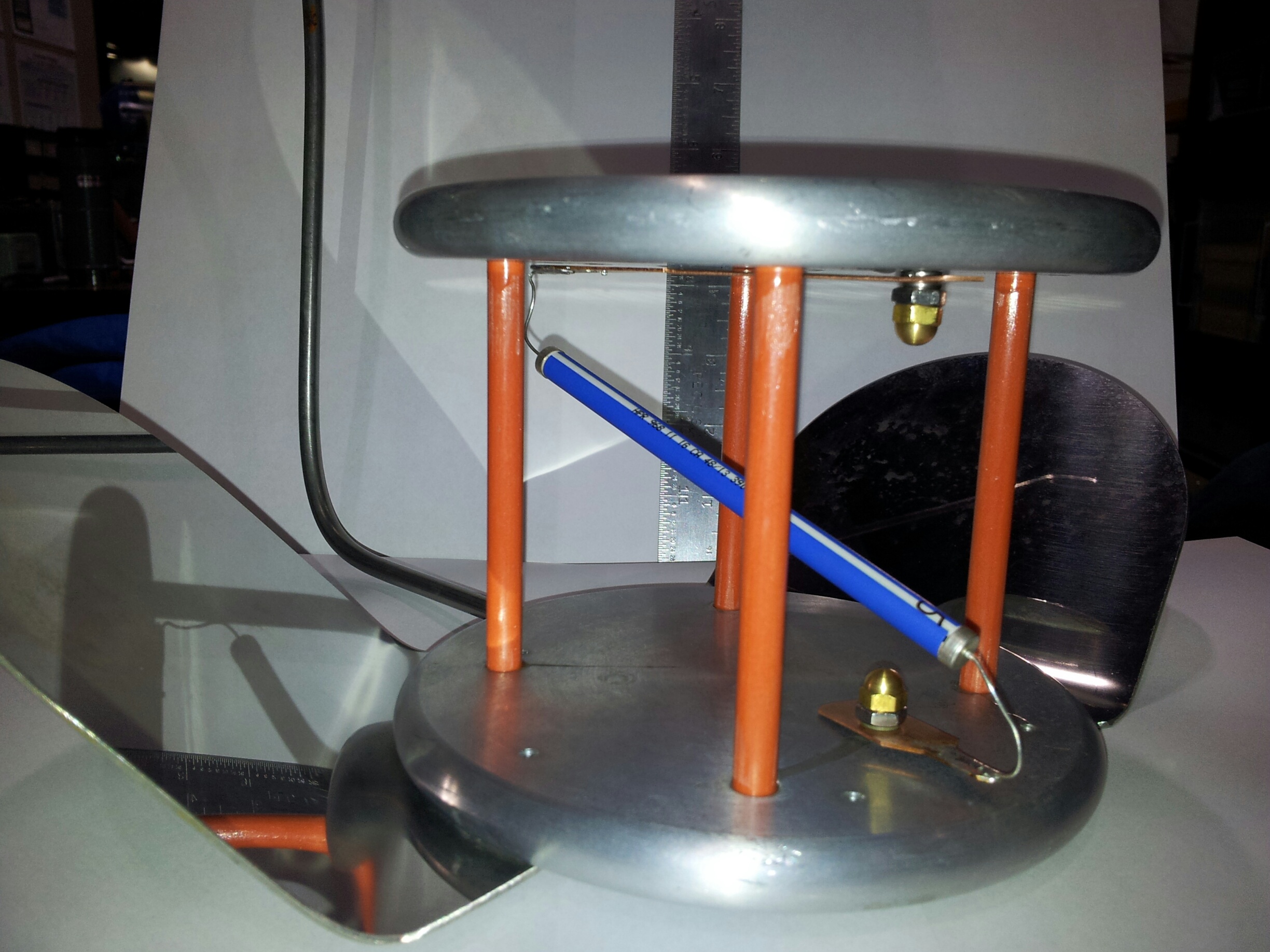}
\caption{A photo of the test apparatus that held the resistors under test.}
\label{fig:apparatusPicture}
\end{figure}

The test apparatus holds the resistors under test and allows the application of the voltage. It is made of two 14.5~cm diameter, 1~cm thick aluminium disks connected by four G10 rods making a total height of 11.5~cm. A hole in each disk aided the mounting of the resistors and provided good electrical contact. The grounded bottom plate had three mirror-finish stainless steel wings attached that were angled so that the resistor under test could be viewed from above. A picture of the test apparatus can be seen in Figure~\ref{fig:apparatusPicture}. Here one of the wings is removed.

\subsection{Method}
We thermally stress-tested the resistors by repeatedly placing them in a bath of liquid argon followed immediately by a bath of lukewarm water.  We observed no cracking of the surface coating by visual inspection, and their resistance, measured at room temperature before and after the thermal cycling, did not change. The resistors were individually mounted in the test apparatus and cooled to liquid argon temperature (88\,K) by immersion in liquid argon. After a short thermalization time, high voltage was applied to the installed resistor. The voltage was controlled by a graphical user interface (GUI) developed for the remote control operation of the Glasmann HV power supply. 

The GUI has two operation modes, either ramping to a fixed voltage (DC), or an ``AC pulse mode'', which ramps the HV quickly to the desired value, confirms the voltage of this setting and than ramps the voltage to 0. The ramp-up time was about 0.5~s, and the duration of the pulse was measured to be about 2.7\,s. The ramp-up time and the pulse durations were the same for all applied voltages.

As shown in Figure \ref{fig:apparatusPicture}, the setup has a capacitance due to its design and the high voltage cabling. During the voltage ramping process, a short and large current spike can be observed, similarly during the ramp down. In DC mode the current is stable and can be extracted for each voltage, in the AC mode the current is changing with the applied voltage, which can be measured only from the internal meters of the Glasmann Power Supply. As the supply is intended for a DC use, the time constant and integration time during ramping does not allow the current to be measured precisely. During the AC pulse, the average peak current value was three to four times larger than the value obtained in the DC mode. This is expected due to the charge and discharge of the setup capacitances in an AC pulse, as a crosscheck the empty setup was pulsed and the same current spike was observed.

A test cycle for each resistor consists of AC HV pulse test, with increments of 2~kV in the applied voltage until either the resistor fails, or the maximum possible voltage of the setup is reached. At some point in this cycle, the resistor could fail due to overpowering the device, usually accompanied by a spark discharge of the HV to ground across the resistor. The breakdown voltage of the resistors was measured with a sample of up to 10 resistors per type, and each resistor was tested for its initial breakdown value, its subsequent breakdown and the resistance after the breakdown. Summarized in Table \ref{tab:results} are the lowest breakdown voltages for the first and subsequent breakdowns of all tested resistor types as well as the power at breakdown. All of them exceeded their HV rating (specified in air) when tested in liquid argon.

\begin{figure}[t]
\centering
\includegraphics[width=0.7\textwidth]{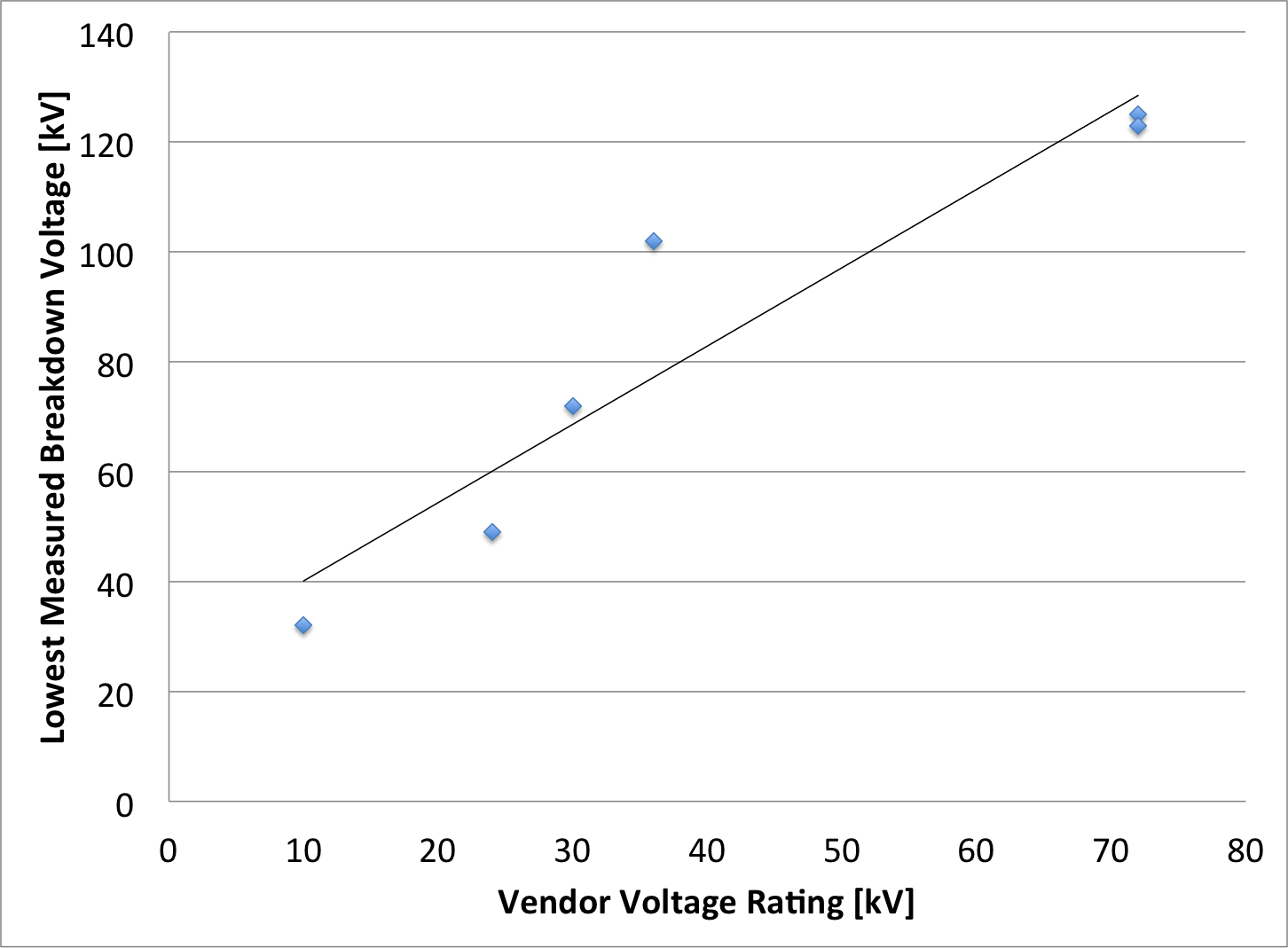}
\caption{Vendor voltage rating in air versus lowest measured breakdown voltage for all failed resistors (i.e. not including the Metallux 969.23).}
\label{fig:vendor}
\end{figure}

All resistor types except for one experienced a breakdown below the power supply voltage limit. Qualitatively the failure voltages follow the power rating for all of the different types of resistors and at various resistances. In Figure~\ref{fig:testohmite} we show the breakdown voltage distribution for one hundred resistors of one resistor type. In Figure~\ref{fig:vendor}, we show the voltage rating by the vendor in air versus the lowest measured breakdown voltage, the dependance of all devices follows a linear fit, the offset can be attributed to the vendors safety factor, which is not known to the authors.

After a breakdown, the voltage of the AC pulse was lowered below the rated breakdown value in air, and again was ramped, until the next breakdown was observed. For all resistors that were not destroyed, the second breakdown happened at lower voltage than the first. For the fatally broken resistors, no second test could be performed.
\begin{table}[b]
\begin{center}
\begin{tabular}{ | p{6cm} | p{2.5cm} | p{2cm} |  p{3cm} |  }
\hline
	Resistor Type and Rating in Air & Breakdown Voltage in kV & Breakdown Power in W & Consecutive Breakdowns in kV\\ \hline
	Ohmite Slim-Mox 104E 1\,G$\Omega$, 10\,kV (air), 1.5W & 32  & 1.0 & terminal damage \\ \hline
	Ohmite Super-Mox 960 1\,G$\Omega$, 72\,kV (air) , 16W & 125 & 15.6 & 51 \\ \hline
	Metallux 969.11 100\,M$\Omega$, 24\,kV (air) , 11W& 49 & 24.0 & 17 \\ \hline
	Metallux 967.15.51 RU 1\,G$\Omega$, 30\,kV (air), 4.5W& 72 & 5.2 & terminal damage \\ \hline
	Metallux 968.10 HPR GD 1\,G$\Omega$, 36\,kV (air), 12.5W & 102 & 10.4 & 83 \\ \hline
	Metallux 968.12 1\,G$\Omega$, 72\,kV (air), 15W & 123 & 15.1 & 83 \\ \hline
	Metallux 969.23 0.5\,G$\Omega$, 48\,kV (air), 23W & >131 &>36.3  & no failure \\
\hline
\end{tabular}
\caption{Result table of the different breakdown voltages. The first breakdown value corresponds to the lowest observed breakdown in an AC pulse for the tested resistor batch, the second breakdown value corresponds to the lowest voltage the resistors were able to withstand after their first breakdown.}\label{tab:results}
\end{center}
\end{table}

For most resistor types, a sample of up to 10 resistors was tested, and a higher statistics sample of the Ohmite Slim-Mox resistors was tested for their breakdown voltage distribution. 

\subsection{Results}
The breakdown voltage of the resistors was measured with a sample of up to 10 resistors per type, and each resistor was tested for its initial breakdown value, its subsequent breakdown and the resistance after the breakdown. Summarized in Table \ref{tab:results} are the lowest breakdown voltages for the first and subsequent breakdowns of all tested resistor types. All of them exceeded their HV rating (specified in air) when tested in liquid argon. 

\begin{figure}[t]
\centering
\includegraphics[width=0.7\textwidth]{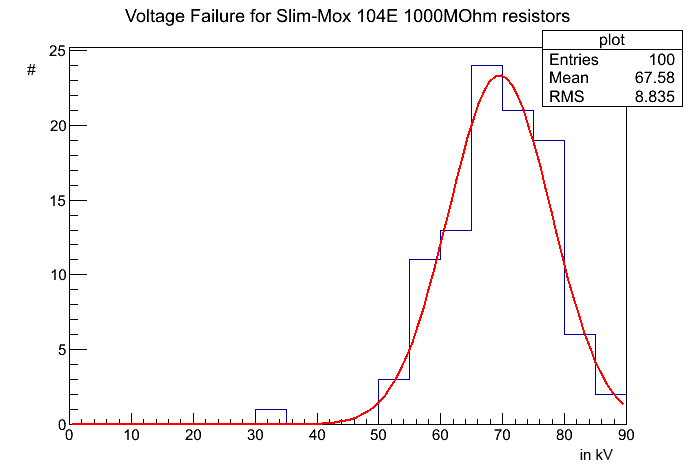}
\caption{Breakdown voltage of the Slim-Mox 104E resistors in a 3\,s voltage pulse.}
\label{fig:testohmite}
\end{figure}

All resistor types except for one experienced a breakdown below the power supply voltage limit. These failures were due to exceeding the vendors maximal guaranteed power rating, i.e. they were expected to fail after crossing that point, though the exact voltage where the resistor failed varies from resistor to resistor and depends somewhat on the details of the testing arrangement.  Qualitatively the failure voltages follow the power rating for all of the different types of resistors and at various resistances. The failure pattern of the resistors can be divided into three categories. First, the flat resistor types Ohmite Slim-Mox and Metallux 967.15.51 were destroyed by the HV pulse. Second, the Ohmite Super-Mox 960 and Metallux 969.11 and 968.10 resistors had large areas of damaged glass coating on the anode side of the resistor, which exposed the inner conductive layers and thus triggered discharges along the surface at lower voltages in subsequent breakdowns. Depending on the pattern of the first damage, the second breakdown happened as low as half the previous voltage rating with a large spread among the resistors. No recovery effect was observed for these resistors, i.e.~the damage was permanent, but stable in subsequent AC pulses. 

\begin{figure}[t]
\centering
\includegraphics[width=0.7\textwidth]{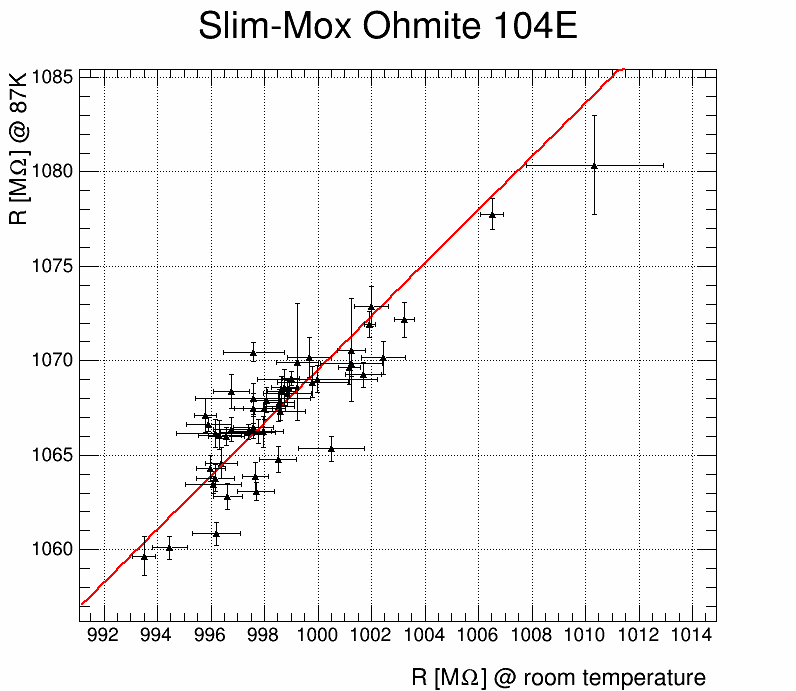}
\caption{Resistance of Ohmite Slim-Mox 104E (1000\,M$\Omega$) resistors measured at room temperature versus at liquid argon temperature of 87\,K. The fit parameters are: $\chi^2$/ndf: $99.09/48$, slope: $1.41\pm 0.01$ and intercept: $-340.7\pm 12.2$.}
\label{fig:testslimmox}
\end{figure}

\begin{figure}[t]
\centering
\includegraphics[width=0.7\textwidth]{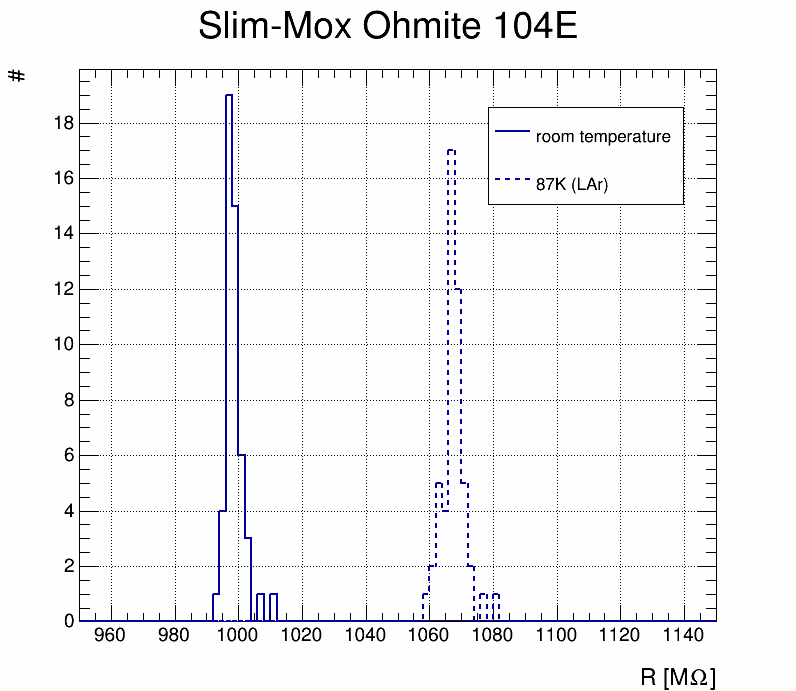}
\caption{Resistance of Ohmite Slim-Mox 104E (1000\,M$\Omega$) resistors measured at room temperature ($998.7 \pm 2.7$ M$\Omega$) and liquid argon temperature of 87\,K ($1067.6 \pm 3.7$ M$\Omega$). }
\label{fig:testslimmox-1D}
\end{figure}

The third type of failure affects the Metallux 968.12 resistors. This resistor has a polymer coating that improves the HV rating in air (and oil) compared with the glass coating. In a breakdown, this coating received less damage compared to a glass coating, and only spark traces along the surface are visible. A small hole in the coating near the anode end of the resistor is the only damage seen. All 968.12 resistors show a similar lowered secondary breakdown voltage with a distribution that is much smaller than resistors with glass coatings. In this case, the subsequent breakdown voltage was lower but remained stable upon further testing. In contrast to the resistors that failed, the Metallux 969.23 resistor does not fail and in fact holds the applied voltages in subsequent pulses. Given the resistance of the filter pot and the resistance of the device and the maximum applied high voltage, this implies a breakdown voltage higher than $(130\pm3)$\,kV.

A high statistics sample of Ohmite Slim-Mox 104E resistors was tested to establish a reliable lower breakdown voltage for this device. Their power rating suggests a failure at a pulse of about 60~kV, yet lower values were observed. In a sample of 100 resistors, the lowest observed breakdown occurred at 32~kV, with a Gaussian distribution around a mean value of 67.5~kV. This measurement, as shown in Figure~\ref{fig:testohmite}, suggests that some outliers can occur, but gives confidence that this type of resistor is able to achieve its nominal 10~kV (in air) rating when used in liquid argon.

\section{Resistance Temperature Dependance}
\subsection{Test setup}
We measured the resistance for both the Slim-Mox Ohmite 104E (1000\,M$\Omega$) and the Metallux 969.23 (500\,M$\Omega$) resistors at room temperature and in liquid argon. This was done using a Keithley 2410 source meter while applying a voltage across the resistor and measuring the current. The resistance of a resistor was obtained by measuring the current at ten different voltages between 100\,V and 200\,V and averaging the results, which also allows the estimation of an error. Fifty resistors were characterized for both models. The test setup has not been calibrated to measure absolute resistance values. 

\subsection{Results}

\begin{figure}[t]
\centering
\includegraphics[width=0.7\textwidth]{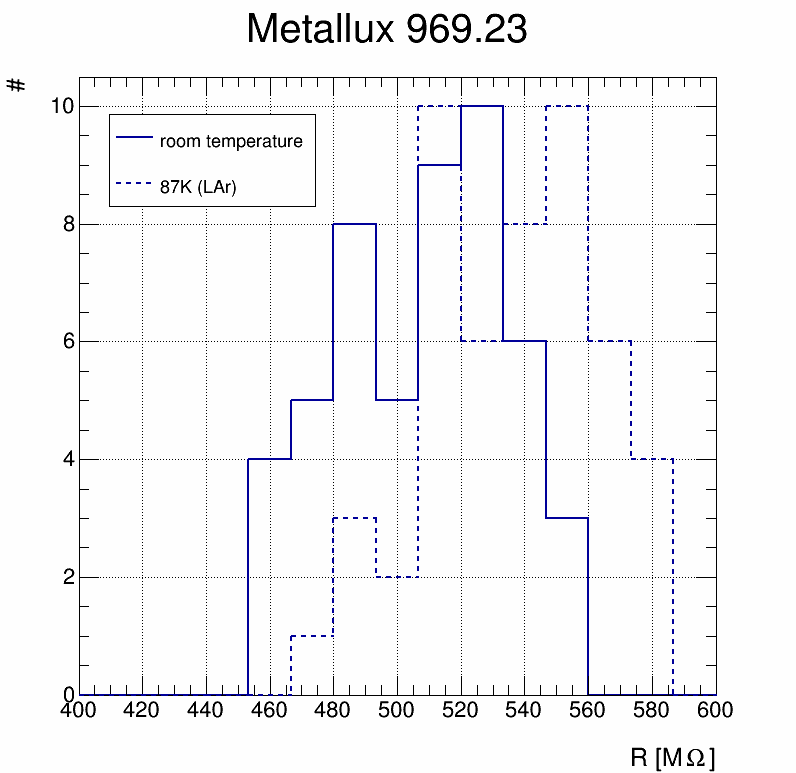}
\caption{Resistance of Metallux 969.23 (500\,M$\Omega$) resistors measured at room temperature ($507.4\pm25.6$ M$\Omega$) and at liquid argon temperature of 87\,K ($536.2 \pm 27$ M$\Omega$).}
\label{fig:testmetallux-1D}
\end{figure}

The mean resistance for the Metallux 969.23 at room temperature is $507.4\pm25.6$ M$\Omega$; for the Slim-Mox Ohmite 104E it is $998.7 \pm 2.7$ M$\Omega$ (see Figures \ref{fig:testslimmox-1D} and \ref{fig:testmetallux-1D}). The smaller spread of the Slim-Mox Ohmite 104E reflects that these resistors were rated with a tolerance of 1\%, while the Metallux 969.23 only have a tolerance of 10\%.

\begin{figure}[t]
\centering
\includegraphics[width=0.7\textwidth]{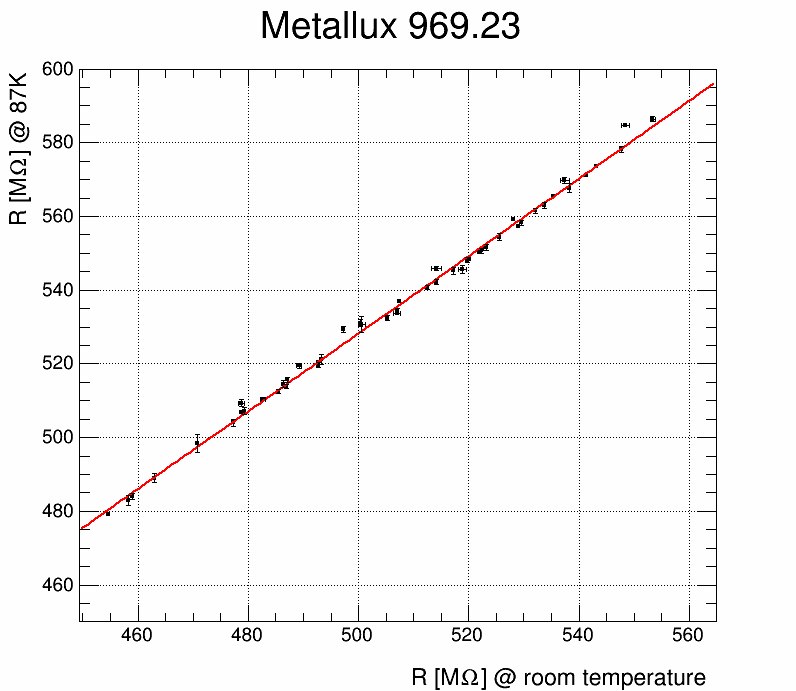}
\caption{Resistance of Metallux 969.23 (500\,M$\Omega$) resistors measured at room temperature versus at liquid argon temperature of 87\,K. The fit parameters are: $\chi^2$/ndf: $199.3/48$, slope: $1.053 \pm 0.004,$ and intercept: $1.714 \pm 1.831$.}
\label{fig:testmetallux}
\end{figure}

It was found that the resistances in liquid argon are higher than the values measured at room temperature. The resistances at room temperature and at 87\,K are correlated as shown in Figure \ref{fig:testslimmox} and \ref{fig:testmetallux}. The measurement uncertainties are due to resolution of the current measurement, with currents in the 0.1 $\mu$A range for G$\Omega$ resistors. 

The mean resistance in liquid argon is $536.2 \pm 27$ M$\Omega$ for the Metallux 969.23 and $1067.6 \pm 3.7$ M$\Omega$ for the Slim-Mox Ohmite 104E. These values are 5.7\% (Metallux) and 6.9\% (Slim-Mox) higher than the resistances at room temperature. The dependence between the value in warm and cold is observed to be linear in the temperature range of this measurement. This scaling behavior needs to be taken into account when using resistors in liquid argon or combining different resistor models in liquid argon applications. 

\section{Summary \& Conclusions}

We tested different types of metal-oxide resistors for their use in liquid argon and found that they are quite suitable for high voltage use in liquid argon. While all resistors survive the cool-down, the flat types are destroyed at extreme over-voltage, while other types receive only surface damage or stay intact. In subsequent breakdowns, the damaged resistor can only hold off lower voltage, but the newer lower breakdown voltage is stable. Different damage patterns and coating materials can explain some of the variations seen between different resistor types. All resistor types were able to hold voltages greater than their ratings in air. The over-power or over-voltage breakdown is correlated, but better than the manufacturer voltage rating of the device. One resistor was able to withstand the maximal applied voltage of the HV power supply. The resistance at room temperature is about 6\% lower than in liquid argon. The resistances at room temperature and in liquid argon are very well correlated so a predictable matching of resistances of different resistor types in a HV divider chain can be achieved by using the measured argon temperature versus air temperature resistance curves.

\acknowledgments
We thank Fermilab for providing funding for the liquid argon and the infrastructure for this measurement. Fermilab is operated by Fermi
Research Alliance, LLC under Contract No. De-AC02-07CH11359 with the United States Department of Energy. We thank LHEP Bern for providing the Metallux resistors.

\end{document}